\documentclass[12pt]{article}

\usepackage[T1]{fontenc}
\usepackage[utf8]{inputenc}
\usepackage{graphicx}
\usepackage{geometry}
\usepackage{color}
\usepackage{xcolor}
\usepackage{hyperref}
\usepackage{array}
\usepackage{amsfonts}
\usepackage{framed}
\usepackage{authblk}
\usepackage{wrapfig}
\hypersetup{
colorlinks,
linkcolor={red!50!black},
citecolor={blue!50!black},
urlcolor={blue!80!black}
}
\definecolor{shadecolor}{gray}{0.9}
\newcommand{\HRule}{\rule{\linewidth}{0.5mm}}
\definecolor{my_color}{RGB}{255,127,0}

\begin{document}
\begin{flushleft}
\begin{tabular}{m{10cm}m{5cm}}
\includegraphics[scale=0.5]{Logo1.pdf} & \hfill \textit{Essay and Opinion
}
\end{tabular}
\end{flushleft}
\begin{center}
\HRule \\ [0.1cm]
{\LARGE \bfseries Novel processes and metrics for a scientific evaluation rooted in the principles of science
\\ \vspace{0.2cm} Version 1
} \\
\HRule \\ [0.5cm]
{\Large \mbox{Michaël Bon\,$^{1}$}, \mbox{Michael Taylor\,$^{2}$} and \mbox{Gary S McDowell\,$^{3,4}$}} \\ 
\end{center}
\begin{flushleft}
\begin{tabular}{l p{0.9\linewidth}}
1. & SJS -- The Self-Journals of Science \\
2. & Department of Physics – Aristotle University of Thessaloniki \\
3. & ManyLabs (www.manylabs.org) \\
4. & The Future of Research, Inc. \\
\end{tabular}
\\ \vspace{0.6cm}
\textit{Made public on Jan, 26th 2017
 under Creative Commons 4.0 Attribution License} \\
\textit{Reviewed and discussed at \href{http://www.sjscience.org/article?id=580
}{http://www.sjscience.org/article?id=580
}} \\
\vspace{0.2cm}
\begin{shaded}
\begin{minipage}{\linewidth} \textbf{Abstract} Scientific evaluation is a determinant of how scientists, institutions and funders behave, and as such is a key element in the making of science. In this article, we propose an alternative to the current norm of evaluating research with journal rank. Following a well-defined notion of scientific value, we introduce qualitative processes that can also be quantified and give rise to meaningful and easy-to-use article-level metrics. In our approach, the goal of a scientist is transformed from convincing an editorial board through a vertical process to convincing peers through an horizontal one. We argue that such an evaluation system naturally provides the incentives and logic needed to constantly promote quality, reproducibility, openness and collaboration in science. The system is legally and technically feasible and can gradually lead to the self-organized reappropriation of the scientific process by the scholarly community  and its institutions. We propose an implementation of our evaluation system with the platform ``\href{www.sjscience.org}{\textit{the Self-Journals of Science}}'' (www.sjscience.org)
\end{minipage}
\end{shaded}
\end{flushleft} 
\newpage

\tableofcontents
\newpage
\section{Introduction: the inherent shortcomings of an asymmetric evaluation system}

Criticism of the current academic evaluation system traditionally focuses on the problematic use of journal reputation as a proxy for scientific quality. However, the harm caused by the research community's dependency on academic journals is more unsettling and destabilizing than we usually think. Journal-based evaluation creates an asymmetry within the scientific community between a minority of scientists sitting in editorial boards, who have the exclusive power to give value to a scientific article by accepting it in their journal, and the vast majority of scholars who strive to convince editors that their articles are highly citable in order to secure one of the limited publication slots, vital for academic career advancement.

In this system of value creation, scientific recognition is artificially turned into a resource of predetermined scarcity for which scholars have to compete. In one camp, members of the scientific community must compete for limited space in a few ``top'' journals, which can impede the natural unrestricted progress of science by disincentivizing open research and collaboration. In the other camp, a low number of editors must also contend with each other for exclusive content to increase the reputation of their journal, a process that can have strong negative effects on scientific output and on the research enterprise as a whole. Although many scholars wear both hats --being authors and journal editors at the same time-- here we do not identify the problem in individual agents but rather in the roles themselves and the power relationship between them. Thus, we argue that it is not only the kind of value that is promoted by the current system that is questionable (journal prestige and `impact', as in \textit{impact factor}): more importantly, it is the way the system produces value and how its implicit asymmetric power structure is detrimental to scientific progress. This fundamental problem must be addressed by any proposed alternative.

In the rest of this introduction we highlight some of the most important consequences of this asymmetry.

\subsection{Peer-trials undermine scientific peer-review.}


Peer-review is a founding principle of science. It is the process through which the community debates over the validity of a scientific proposition. It allows science to be self-correcting and to develop beyond the prejudices of the few.

In the current publishing environment, since scientists are competing for the same limited resources, relations between peers can become inherently conflictive. For instance, scientists working on the same topic may tend to avoid each other for as long as possible so as not to be scooped by a competitor, whereas collectively it is likely that they would have benefited most from mutual interaction during the early research stages. The most worrying consequence of peers' diverging interests is that debating becomes socially difficult --if not impossible-- in the context of a journal. The rejection and downgrade of an article to a lower-ranked journal can be a direct consequence of a scientific disagreement that few people would openly take responsibility for, to avoid reprisals. 
 
While the reliability of science comes from its verifiability, today it is being validated by a process which lacks this very property. Journal's peer-review is not a community-wide debate but a gatekeeping process tied to the local policy of an editorial board, where a small number of people hold temporary authority over an article, and whose goal is to support a binary decision or acceptance of rejection within some deadlines.  We propose to rather refer to journal's peer-review as a `\textbf{peer-trial}', a term that in our opinion better accounts for its goal and the social dynamics at work behind it. Since peer-trials necessarily involve a degree of confidentiality and secrecy, and since they are limited in time, many errors, biases and conflicts of interest \cite{randomPR, resistance,ceci,PR_female} may arise  without the possibility of correction. In that sense, peer-trials are questionable as a scientific process.

The peer-trial has become the standard in scientific publishing for the past 60 years \cite{mnielsen,PR_history}, and is the only modus operandi that the current generation of scientists has known, entrenching the belief that passing them is equivalent to attaining scientific validity. This is not to say that they are unfair, unuseful, or without intellectual added-value when all participants are competent, genuinely committed and have the interests of scientific truth at heart. Surveys report that 90\% of authors, whose article has been accepted, feel that peer-trial had improved it \cite{PRsurvey, PRsurvey2}. Nevertheless, this is not a guarantee of scientific validity. Limitations in time and the insufficiency of available competences mean that a perceived improvement does not necessarily achieve high scientific standards \cite{error} as expressed, for instance, by the general misuse of statistics in biomedical sciences \cite{false,pvalue}.

Moreover, improving scientific validity after a peer-trial process is quite different from evaluating scientific innovation. Articles that may best contribute to the progress of science are often unexpected or disruptive to the status quo  \cite{sci_rev}. It is precisely in this context that the peer-trial format is most likely to fail and go wrong \cite{horrobin}, with unverifiable shortfalls for science.

While peer-trial still dominates the mainstream, there are strong signs that the scientific community is actively engaged in a more continuous process of validation. Browsing websites such as \href{https://pubpeer.com}{PubPeer} or \href{http://home.publons.com}{Publons} (where ``post-print peer-review'' is possible) makes it clear that, although articles are improved with respect to initial submission, the discussion process continues long after publication and that the evolution of articles is a more dynamic construct \cite{dynamic_article}. This is at odds with the world of undisclosed email dialogues between authors and editors, and reviewers and editors during the peer-trial process.

The unaccountability of peer-trials have further systemic consequences. For instance, referees cannot be credited for their work and institutions are led to promote a one-dimensional definition of scientists' utility, that would only rely on their productivity, and whose latest avatar is the h-index \cite{hindex}.

\subsection{Competition between journals excludes research with perceived low impact.}
Editors cannot afford to neglect the impact factor of their journal \cite{impact_game}. Instead, they are motivated to inflate it by selecting articles that they think will be highly cited in the following two years. This selection bias prioritizes research that is more likely to trend at the expense of elements that are critical in the testing of scientific ideas but not conducive to the increase of a journal's impact factor (such as high-accuracy experimental data that did not prove a ``positive'' effect, or replication studies).

\subsection{Gaming the system results in low quality.}
Since the value of an article at present is tied to being published in a journal, and since peer-trials are not transparent, can be biased, are few in number and highly variable in quality, scientists may be tempted to game the traditional publication process. By gaming, we mean that a scientist may inappropriately generalize from the trivial, avoid statistical rigour, mislead by an elegant narrative, exploit power in the relationship between editors, reviewers and authors \cite{politics}, or even use fake identities \cite{fake_ID} and commit fraud\footnote{The latter can be followed on \href{retractionwatch.com}{Retraction Watch}, a blog which keeps an active record of retraction cases, and investigate many.} \cite{fraud}. For instance, statistics from the Nuffield Council on Bioethics on the culture of scientific research in the UK show that 58\% of survey respondents reported that they were aware of scientists feeling tempted/under pressure to compromise on research integrity and standards, and one-third of scientists under 35 reported feeling this pressure themselves \cite{fraudculture}. The major consequence of gaming is loss of quality which leads to irreproducible research getting the seal of approval by the publishing system \cite{irr_cancer, irr_psy, irr_eco}.

Once an article has been published, the editor may be reluctant to have a debate open, whose outcome may damage the reputation of the journal \cite{raphaPNAS,noopenreview}. The `time-to-retraction' (i.e. the time from publication of an article to publication of retraction) averages 32.91 months \cite{retraction_time}. This is a concern for debate in science, especially in the sphere of public health where clinical trials can already have reached an advanced stage by this time \cite{trial}. There is a fundamental contradiction between the scientific need to constantly and dynamically debate, test, refine or correct scientific claims, and the private need of third parties to deliver and sell something as a static end-product.

All of this can considerably delay or hinder the self-correction of science, result in a waste of time and (public) money, and can have deleterious effects on the credibility of science and what it produces.

\subsection{Scientific conservatism is placing a brake on the pace of change.}
It is questionable whether any static subset of the scientific community could appropriately manage the direction of science. Today's highly-regarded researchers naturally tend to defend the paradigm that underpins their reputation, whilst opposing tomorrow's ideas \cite{funeral}. 

For example, in ``The Dynamic State of Body Constituents'' \cite{schoenheimer}, Schoenheimer tried to introduce the concept that proteins were broken down intracellularly (a fact that we now take for granted, with the study of ubiquitin). However, it took more than 30 years for this concept to be accepted, and likely delayed the discovery of ubiquitin, in part because Nobel Laureate Jacques Monod was a proponent of the theory that proteins, once formed, were ever present \cite{ubiquitin}.

In science, good ideas may eventually prevail, but a lot of time and effort may be mis-spent before they do. Though the beauty of the scientific method is to allow humans to go beyond their own prejudices, the traditional publishing system is prone to working in favor of the current dogma. The main mechanism of selection of editors (i.e. co-opting between reputable researchers) is creating and enforcing additional constraints and limits on the progress of science without the security of collective wisdom. \\

To address these issues, in this article we introduce a model based on a novel, open, and community-wide evaluation system that captures a well-defined notion of scientific value and which is based on scientists' collective intelligence and judgement. We advocate that the inherent logic of our model may reverse the process of privatization and fragmentation of evaluation associated with the use of journal rankings. Moreover, the system is technically and legally feasible in the current environment. In Section \ref{sec2} we define scientific value, the methodology used to assess it and associated metrics. In Section \ref{sec3} we discuss the mechanics and merits of our proposed model. In Section \ref{sec4} we highlight some implications of this novel way of evaluating research works in the context of a global competition for money, tenure and honors. Our treatise is part of a broader vision that is also developed in \cite{Moi} and \cite{interview}.

\section{A symmetric process for the creation of scientific value} \label{sec2}

In this section, we present a definition of scientific value and describe the open and community-wide processes required to capture it. These processes maintain symmetry in the creation of scientific value and fulfil what we consider the minimal expectations from any desirable alternative evaluation system, which are:

\begin{enumerate}
\item to promote scientific quality.
\item to provide incentives to authors, reviewers and evaluators.
\item to promote academic collaboration instead of competition.
\item to be able to develop in parallel to current journal publication practices (as long these remain essential for funding and career advancement).
\item to propose article-level metrics that are easy to calculate and interpret.
\item to be verifiable and hard to game.

\end{enumerate}

A prototype of an evaluation system driven by these processes is implemented in ``the Self-Journals of Science'' (SJS, www.sjscience.org): an open, free and multidisciplinary platform that empowers scientists to achieve the creation of scientific value. SJS is a horizontal environment for scientific assessment and communication and is technically governed by an international organisation\footnote{\url{www.openscholar.org.uk}} of volunteer research scholars whose membership is free and open to the entire scientific community.

\subsection{Scientific value as validity and importance}

A scientific article relies on refutable statements that contribute to a body of knowledge about the object of study. In this description of science, the value of an article spans two distinct notions, that require their own assessment mechanisms: the correctness of its statements, which we call its \textbf{validity}, and the value of its contribution, which we call its \textbf{importance}.

The validity of an article is established by a process of open and objective debate by the whole community. Since the contribution of a scientific article essentially relies on refutable statements, debating them in principle \footnote{that is, in the limit of infinite time and infinite resources to test the statements} can eventually converge on a consensus about whether it has reached accepted scientific standards (and what these standards should be: methodological soundness, unambiguity of of presentation, satisfaction of various protocols, inclusion of appropriate references, etc.) or whether or not it needs further revision(s).

The importance of an article is the outcome of its perceived importance by each member of the research community. This perceived importance is a subjective assessment that depends on personal knowledge and understanding, intuition, and anticipation of future advances in the field. Unlike validity, the perception of importance does not rely on refutable elements that could be used to automatically resolve disagreements. Broad consensus is not expected based on the importance of an article alone; for instance, scientists might rightfully diverge in their belief that a certain path of development of their field is more valuable than another. The importance of an article is thus socially determined and field, time and culture-dependent.

The two different concepts cannot be measured by a single index: a methodologically valid article may not be important just as a trending article may prove to be wrong. In the traditional system, these notions are merged and uniquely expressed by a local, opaque and one-time event: journal publication.\footnote{In our implementation, we also introduce an addition notion \textit{priority} which is of practical use but not an index of quality of a scientific article. Nonetheless, we have introduced priority as a simple counter that allows scientists to mark articles prior to thorough scrutiny in the same vein as the ``Like'' button on Facebook. Priority is a short-term notion that does not imply a scientific judgement and the reasons for prioritizing an article can be either positive (e.g. the article looks interesting) or negative (it looks terrible and needs a quick refutation). Priority offers a instant filter to help organize scientific output on a short-time scale, before assessment of its validity and importance. The reason this is important is because it allows the system to channel misleading short-term effects (such as those of a promising abstract combined with an overstating narrative) into an index that is different from that used for scientific evaluation.}

\subsection{Assessing validity by open peer review}

We have defined scientific peer-review as the \textbf{community-wide} debate through which scientists aim to \textbf{agree} on the \textbf{validity} of a scientific item.

In our system, peer-review is an open and horizontal (i.e. a non-authoritative and unmediated) debate between peers where ``open'' means transparent (i.e., signed), open access (i.e., reviewer assessments are made public), non-exclusive (i.e., open to all scholars), and open in time (i.e. immediate but also continuous). This brings a new ethic to publishing \cite{Moi}: the goal of peer-review is not to provide a one-time certification expressed in the form of a binary decision of accept or reject as per the traditional mode of publishing, rather it is to scientifically debate the validity of an article with the aim of reaching an observable and stable degree of consensus. Here, reviews are no longer authoritative mandates to revise an article, but elements of a debate where peers are also equals. The influence of a review over an article is based on its relevance or its ability to rally collective opinion, and on an open context where authors cannot afford to let relevant criticism go unanswered.

The validity of an article is captured by a transparent and community-wide vote between two options: ``this article has reached scientific standards'' \footnote{The notion of ``scientific standard'' is not enforced by any authority but is dynamically defined by the majority vote that follows a global conversation about each article.} or ``this article still needs revisions''. \textit{The quantifiers of the validity of an article (i.e. its metrics) are hence the number of scientists who voted, and the fraction who validated the article}. Any reader can instantly access the current state of the voting process which is displayed at the header of the article (Figure 1). 

In our proposed implementation of such open peer-review, we introduce other features aimed at incentivizing positive interactions between participants and the proper self-organization and self-regulation of the debate:
\begin{itemize}
\item Articles are interactive and reviews are appropriately embedded into them. Reviews therefore benefit from the same visibility as the article.
\item Reviews can be individually evaluated with a +/- vote system
\item The vote ``this article still needs revisions'' must be substantiated by the writing of a review or up-voting of an existing review.
\item Upon revision, authors can select those reviews most useful to them and the reviewers get proper acknowledgement in all subsequent versions of the article - in their header (online) or in the cover page (PDF).
\item The history of the article is always accessible.
\end{itemize}

In this form of self-publishing, scientific articles become dynamic \cite{dynamic_article}, authors become active in energizing a peer-review process \cite{selfpub} whose quality is driven by collective intelligence in a symmetric environment where the best scientific ideas are subjected to natural selection \cite{naturalselection}. The process produces a text initially written by its authors but which also includes the debate it has generated within the community.

\begin{figure}[h!]
\includegraphics[width=\textwidth]{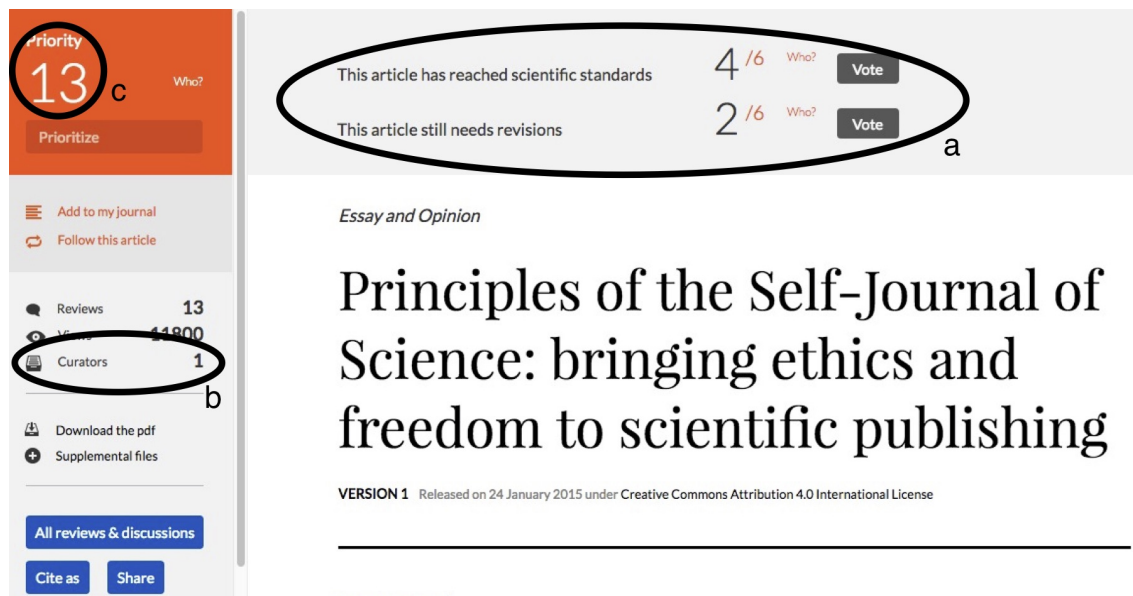}
\caption{\textit{Header of an article uploaded on SJS. It features its validity (a), its curators (b) and its priority (c) (see footnote 4). Each of these metrics is transparent and the identity of scholars who contributed to them can be accessed with the ``Who?'' button}}
\end{figure}

\subsection{Assessing importance with \textit{self-journals}}

\textbf{Importance vs impact.} A fundamental difference between our understanding of importance and the notion of impact as measured through citations (e.g., the impact factor, the journal citation distribution \cite{cite_dist} or the RCR \cite{RCR}) or usage statistics (e.g. number of tweets, hits, downloads, etc), is that importance should be explicitly and directly related to a scientist's judgement. In that sense, metrics of impact do not necessarily convey useful or separable information about importance.

As for citation-based metrics, we argue that a citation is not an endorsement of the article we cite. For instance, we will cite an article we may want to scientifically refute; citations may be included so as to please an anonymous referee \cite{cite_reviewer}; we often cite review articles instead of original articles, or we simply copy and paste or incorrectly cite without reading the actual articles \cite{cite_without_read}. Also, the narrow context of our own publications may \textit{not} be relevant to cite and give credit to articles that have, however, been important to us. 
 
Usage statistics (altmetrics) are even less reliable since they can be massively gamed and may be derived from sources unrelated to science, or even from a plethora of ghost accounts on social media. \\

\textbf{Self-Journals.} In our alternative evaluation system we introduce the concept of \textit{self-journals} as a way for scientists to properly express their judgement regarding an article's importance for a specific field. A self-journal is a novel means of personal scientific communication; it can be thought of as a scholarly journal attached to each individual scientist that works on the curation of any scientific item available on the Web via hyperlinks (and not on appropriation of articles following a submission process). 

A self-journal is released into structured \textit{issues}, which are collections of articles around a certain topic. Every issue has its own title and editorial providing an introduction for the community and must contain a minimal number of articles (in our implementation, we set this minimum to 4). The curator has the possibility to provide personal comments on each article that has been curated in the issue (for concrete examples, please check the first issue of the self-journal of \href{http://www.sjscience.org/memberPage?uId=148&jId=10#journal}{Sanli Faez}, \href{http://www.sjscience.org/memberPage?uId=90&jId=6#journal}{Konrad Hinsen} or \href{http://www.sjscience.org/memberPage?uId=1&jId=14#journal}{Micha\"el Bon)}. 
The consistency of the selection of articles and the relevance of the personal comments determine the scientific added value of each self-journal issue. 
Every scientist can curate their own self-journal, through which they can release as many issues on as may topics as they please. Curators can take advantage of self-journals to review a field, present a promising way to develop it, offer a comprehensive collection of their own research,  host the proceedings of a workshop or a journal club, or popularize scientific ideas and discoveries etc. A self-journal reflects the scientific vision of its curator and bears his or her signature. Interested readers can freely subscribe to a self-journal and get notified whenever a new issue is released.

\textit{An ecosystem of self-journals offers a way to quantify the importance of an article, primarily by the number of its curators}. 

\section{Benefits of this evaluation system} \label{sec3}

The evaluation system we propose generates easy-to-use article-level metrics for validity (the number of scholars engaged in peer-review and the fraction of them who consider that the article is up to scientific standards) and for importance (the number of scholars who have curated the article in their self-journal). Such metrics can progress in parallel to existing publication practices and metrics. On the one hand, the assessment of importance through curation via  hyperlinks can apply immediately to all scientific literature on the Web, irrespective of its legal owners. On the other hand, the assessment of validity requires only the uploading of author-copyrighted content or pre-prints -- a practice which is already standard in some fields such as physics, and gaining momentum in others such as biology \cite{bio_preprint} (see for instance \href{http://asapbio.org}{ASAPBio}). Below we discuss how this evaluation system satisfies the properties that we laid out in section \ref{sec2} and the benefits it provides.

\subsection{Avoiding conflicts in evaluation}

\textbf{Competition for an abundant resource promotes actively collaborative behaviour.} In our evaluation system, the conflict-creating competition for rare slots in a top journal is replaced by a competition for open peer recognition which is \textit{abundant} and \textit{non-exclusive}. Giving recognition to a peer does not deprive somebody else of it; an important article does not rise at the expenses of any other article. In this symmetric recognition-based economy of knowledge where scholars are at the same time authors, reviewers and evaluators of science, all scholars can give value to their peers and are potential benefactors of each other. To get the attention of peers and convince them of your scientific merits, the best \textit{strategy} (in the sense of game theory) is to pay attention to their work and rightfully give them the recognition they deserve in a way which lets your own expertise publicly shine. This is exactly what is achieved by self-journals and open peer review.  \\

\textbf{Convergent interests in open peer-review.} Peer-review is transformed from an authoritative trial into an open, transparent and unmediated scientific conversation. It does not assume which participants (either authors or reviewers) will be right in advance and does not result in a definitive and binary decision of acceptance or rejection of an article. Then, openly disagreeing on a constructive basis becomes an opportunity for both the author and the reviewer to gain recognition, both from each other and from the rest of the community.
\begin{itemize}
\item For authors, an ongoing expert debate raises the level of awareness of their article among the community. A spontaneous, signed and relevant critique is delighting evidence that a peer cared and gave some thought to their article. The reviewer may eventually turn into a curator once appropriate answers have been given. A live process of improvement of the work under review is also the best way to convince other readers that articles and research results are valid.

\item For reviewers, credit is received for their public display of expertise, enhancing their reputation within the community. Their reviews become an integral part of the paper and are visible in all its instances and they directly benefit from helping raise the visibility of the article they have chosen to review. In this way, reviewers get special credit and are incentivized to be relevant, constructive and courteous. The public display of expertise by engaging authors in discussion will prove beneficial to reviewers at a later stage when they author their own papers, as their reputation within the community is enhanced also by the quality of their ability to review.

\end{itemize}

\textbf{No conflict in the assessment of importance.} The assessment of importance is also devoid of any conflicts. A self-journal is a personal tool whose content, topics of interest and release dates are completely up to its curator and his or her private interests, policy and communication strategy. There is no submission process; articles are freely picked by the curators according to the message they would like to deliver to the community in the next issue of their self-journal. The consequence of this is that no single article can be expected to be curated by a specific self-journal. An unimportant article is simply characterized by the fact that it remains in its default state with no curator, and not by any explicit and conflictive statement that it is unimportant coming from somebody in particular. The act of assessment of importance by openly taking responsibility to include such an article in an issue of one's own self-journal, can only be positive by mutually increasing the visibility of both.

\subsection{Incentives for maintaining one's self-journal}

\textbf{Incentives in the absence of official recognition by institutions and funders.} Self-journals have their own rationale. Firstly, they are a means of personal scientific communication that allow their curators to elaborate on an individual vision of science with the necessary depth they see fit. A self-journal therefore provides a great scientific service to its readers by providing a level of consistency in the interpretation and analysis of scientific output. In return, this service benefits curators who increase their visibility and influence over what is disseminated to the community.  Self-journals give new freedom and scope to the editing process since curation, as proposed here, applies to any research work with an Internet reference. In other words, a mechanism is provided that allows scientists to fully express an aspect of their individual worth that is absent in the current system, and build a reputation accordingly.

A response is also provided to the problem of the decreasing visibility of authors, articles and reviewers as the volume of scientists and scientific works grows on the Web. Each issue of a self-journal acts as a pole of attraction that is likely to have a minimum audience: the authors whose articles have been curated can be notified about what is being said about their work, and may want to follow the curator. Moreover, on a platform like SJS where the ecosystem of self-journals is well integrated, interest for a particular article can guide readers to self-journal issues where it has been uniquely commented on and contextualized in relation to other articles.

We wish to emphasize that the interest value of a particular self-journal issue does not lie so much in the intrinsic value of the articles selected, but rather in the specific comments and collective perspective that is being given to them. Consequently, if a certain article is curated by a ``reputable'' scientist, its other curators will not lose value or visibility - even when their issues contain exactly the same articles; different self-journal issues can always maintain independent interests. Every scientist therefore has both a short-term and long-term personal interest in maintaining a self-journal, and in reading those of their peers.  

It is also worth mentioning that the expected primary use of a self-journal - sharing one's vision of a field, is not time-consuming as scientists have already developed such visions in the course of their research whenever they have produced a bibliography. Maintenance of a self-journal in this case is reduced to the time it takes to select the most important articles and to write useful comments if desired. \\

\newpage
\textbf{Incentives with official recognition by institutions and funders.} 
If, as an evaluation system, self-journals attract the attention of institutions and funders, then the incentive for maintaining one's own self-journal is clear: it is an indispensable tool for self-promotion that allows everyone to have a share in the direction of evolution of their field. That power, which is today in the hands of editors in a top-down certification system, will be redistributed horizontally amongst scientists who are collectively co-responsible for validation and evaluation and are properly credited for doing it well. Growing awareness of the positive impact of ease-of-access to validated, important-ranked and credible science, as well as the increase in scientific value given to researchers and, by proxy, their institutes and funding bodies, will help generate momentum for the mass adoption of self-journals and open peer-review. In turn, mass adoption is what will bring the power and richness of this evaluation system at its peak. 

\subsection{Promotion of scientific quality} 

\paragraph{The need for quality curation.} To gain followers, readers and influence, scientists should maintain a thoughtful and interesting self-journal. In other words, they are incentivized to be good evaluators, i.e. to carefully select and curate relevant collections of articles in issues and enhance them with enlightening comments. Scientists that mechanically curate all articles in their field (akin to copy-pasting entire bibliographies) without adding value by providing comments or constructive criticism, will receive less attention and influence than one who focusses on quality.

Moreover, because of the need for structure and the writing of an editorial for each issue, self-journals are not automatons. They are not a time feed to which articles are appended in a single click and where the latest additions decrease the visibility of prior additions (something that can occur on very short timescales e.g. minutes), such as in other environments. This ``friction'' ensures that self-journal issues will be released only when a curator has a point to make, and that qualitative strategies are more efficient than quantitative ones when the aim is to get as large audience as possible within the scientific community. 

Thus, unlike simple popularity contests (such as the number of ``Likes'' on Facebook), there are a feedback loop and constraints acting on the quality of the self-journal itself which encourages its curator only to include articles that reach his or her standards.

\paragraph{Curation recognizes works necessary to the progress of science.} The value of a self-journal built on the validity and importance of its articles is driven by the scientific interest and relevance of the vision developed by its curator, and not because it has monopolized the distribution of highly-citable articles as is the current situation. Consequently, as an evaluation system, it liberates articles that are essential to scientific progress which are today penalized because they are believed to be less citable (e.g. replication studies or reports of valid results which failed to prove the author's hypothesis; often unwisely referred to as ``negative'' results). Such contributions are as important to the scientific method as reports of statistically-significant effects, especially in the empirical sciences. When providing their analysis of a certain topic, curators are now free to integrate items they see as important to making their point as convincing as possible.

In the case of replication studies there is actually an incentive for their curation. Scholars who have already curated a particular standard are interested in knowing whether or not a certain claim has been confirmed or refuted. Thus, it is likely that the replication of an original result will benefit from the same level of curation as the original article. This also means that a laboratory which is considering replicating a result can have an \textit{a priori} estimate of the publishing reward expected for investing time in performing an experiment. Results that are regarded as important and highly curated provide a stronger incentive for their replication. This provides a novel positive feedback mechanism and impetus for science to move forward confidently, since major results will be verified.

\subsection{No artificial constraint on time, space and format }

Unlike academic journals, self-journals are not expected to curate only the most recent articles. It is in a curator's best interest to provide a mix of both past and present articles when creating a deep and comprehensive vision of their field \cite{Sanli}. For an article, this implies that its evaluation span becomes time-dependent. For instance, a disruptive innovation which is gradually understood will become increasingly curated and reach high importance even if the author was the only one to understand it at the beginning of the process. 

Furthermore, linked data that results from hyperlinking content frees self-journals from space and format restrictions, present in the current mode of publishing. By this we mean that the content of self-journals is not constrained to any particular type of scientific item (article, thesis, conference proceeding, poster, essay, technical report etc), design format or storage requirement. This mode of evaluation then, gives the scientific community a certain freedom to evolve the format of  articles according to its needs.

\subsection{Robustness to gaming and biases} 
Here we highlight how the inherent logic of this evaluation system and its horizontal power structure, fight against gaming and human biases. However, it is clearly impossible for us to be sure how a culturally diverse population of millions people, subjected to different local constraints, would behave in such an environment. We are therefore looking for as many feedbacks as possible on this topic as well as practical tests.

\textbf{Gaming.} The metrics we propose are established by open and community-wide processes which make them hard to game. For instance, unlike citations and usage statistics, two accomplices are unable to create an infinite loop that boosts the importance of each other's articles. This is because the primary quantifier is the number of curators rather than the number of self-journal issues in which an article appears (which would become analogous to journal citation). If a scientist exploits friendship bias (i.e. they curate articles of an acquaintance multiple times), they would only increase their index by +1. At the same time, their self-journal would lose credibility in the eyes of the community and damage their reputation. Gaming a truly open peer-review and evaluation system like the one proposed here, would imply a successful manipulation of a significant fraction of the scientific community in open and transparent processes or involves the (highly unlikely) \textit{joint} individual misconduct of a large fraction of scientists which is easy to expose as it is large. The difficulty of these scenarios and the ease at which they can backfire strongly dissuade them. 

Note that this evaluation approach does not assume that the majority of members of the scientific community are virtuous either. It is peer pressure which naturally and constantly exerts itself to enforce the highest scientific standards everywhere. In such an environment, the best self-interest of each scientist is aligned with the ethical requirements of science. There is no way to get recognition other than making valuable contributions to scientific knowledge, be it in the form of articles, reviews or a self-journal.

Finally, since these processes of peer-review and curation are not locked up in a proprietary database, the research community is free -- and expected -- to develop open source modules that can detect anomalous behaviors and properly address any gaming scenario. Such modules could be for instance integrated in search engines for signaling, or could be run independently. They will enforce further self-regulation and can demonstrate the reliability of these processes as they gain momentum as an evaluation system.

\textbf{Biases}. Similarly, the absence of vertical relationship between scientists (being replaced with horizontal and reciprocal ones) and the full transparency and accountability of the system, combine to oppose the negative expression of human biases with respect to specific works or fellows (i.e. conflicts of interest, gender-, race-, age-, country-based biases, etc.)
For instance, the fact that reviews are not authoritative and are themselves subjected to peer scrutiny (with the possibility of being evaluated with a +/- voting system and being refuted by the authors when irrelevant) strongly incentivizes reviewers to strengthen valid scientific arguments. Doing otherwise might backfire and negatively impact the reputation of the (always non-anonymous) reviewer. In the logic of our system, relevant reviews contribute to attracting the attention of the community and give life to an article. Therefore, even if a reviewer wants to criticize an article because of a personal bias against the authors, the need to do so in accordance with high scientific standards actually results in a benefit to the authors.

Moreover, every scientist is generally incentivized not to express negative biases against their peers because they are the ones with the freedom and power to endow value to their works. Thus, failing to reach a subgroup of peers for reasons other than science goes against the interests of the scientist, who may suffer from a shortfall in the evaluation of his or her works.

In addition to deflecting or opposing the negative expression of a bias, the system possibly also offers a long-term opportunity for a desirable cultural evolution in relation to disfavored minority groups. The visibility of their individuals no longer depends on a self-reproducing vertical power structure which can limit their contributions (e.g. by making it harder for them to participate in the publishing process as an anonymous reviewer or editor). Rather, they can autonomously and publicly express their value in all dimensions of the scientific activity. This is an opportunity for them to take, that cannot go unseen by the community, and which will help form the mindset of future generations.

\subsection{No loss of information} 

The current ubiquitous use of indices in decision-making by administrative structures and funders enforces the constraint that our evaluation system generates numbers so that it can be easily adopted and recognized. We have therefore proposed article-level metrics for  validity and importance, but it is obvious that  expressing something as complex as scientific value in the form of a few numbers inevitably implies a loss of information. The quantifiers proposed here are mere tools to provide a first sound and reliable picture of what is valid and important in science, and we argue that they are in many respects preferable to journal rankings and impact factors in capturing scientific value.

However, in our evaluation system, the qualitative processes underlying the computation of our metrics remain fully accessible and they offer much richer and accurate information that can and should be made use of. An ecosystem of self-journals provides a context in which an article can be appreciated via its relationship to the other articles of the issues in which it has been curated. The perspective of curators is explicitly expressed in narrative form via  editorials and individual comments. Science can be followed in many additional ways: by following the activity of specific scientists (i.e. what they publish, review and curate) that one considers to have better skills than the rest, by following the activity and connections developing around specific articles, by reading self-journals. Networks of co-curated articles, of authors, of reviewers and curators can be easily studied in a systematic and automatable fashion.  The collective intelligence of the scientific community can be dug as deep as necessary by all users, and according to their needs.  

This complete and easy-to-process information displays a diversity of approaches that opposes an evolution of our evaluation system into another ``tyranny of the metrics'', where the only visible articles would be those which rank at the top according to our quantifiers.

\subsection{Acceleration of the adoption or debunking of novel ideas} 

The traditional publishing system has an inherent conservatism but also the structural means to slow down the adoption of disruptive ideas if not ``believed'' by dominant voices. Such mechanisms exploit the asymmetry of the evaluation system and are absent from the self-journal system. However, self-journals alone are not sufficient to overcome the natural reluctance of a community to adopt novel ideas. There is inertia. Ideas may be expressed by authors' articles and self-journals but lack of peer commentary by the community is a challenge also for the use of collective intelligence for building scientific value. Our argument is that rewardable open peer-review is the motivator and accelerator of such a process. Indeed, the author of a disruptive idea can contribute as a reviewer of peer articles and accordingly point out their possible shortcomings; the community will eventually be triggered into noticing the novel idea and responding - leading either to its more rapid adoption or debunking.

\subsection{Full achievement of Open Science}

The way that self-journals create scientific value produces an additional mechanism  that triggers openness. When the goal of a scientist is shifted to convincing as large a section of  the community as possible (and not simply matching the minimal standards of a journal policy), those who practice open science have an advantage over those who do not. Ensuring that the full text of article, data and code are all freely and openly available, is the vehicle for maximizing the potential for getting positive feedback from the community. If, for instance, data is not made available, members of the community are likely to conclude that the article cannot be validated (since it cannot be tested), and the authors will be penalized. With the evaluation system we propose, total openness is in the self-interest of every scientist and occurs naturally without the need for top-down mandates or bureaucracy. 

We believe that the correct battle for Open Science is the one for the metrics that reward it. Community-based metrics re-empower the average scientist, while at the same time providing incentives to energize and reward collective participation. They will strengthen and reach beyond open access. Paradoxically this battle is also much easier than that for open access since it only depends on scientists and not on what legacy publishers can think or do.

Moreover, since the assessment of validity and importance of articles takes place in a self-organized and self-regulated way, there are no intermediaries between scientists and the direct costs of scholarly communication are reduced to the cost of storage. This is presently tiny in comparison with what is currently being paid by institutions to legacy publishers whether in the form of journal subscriptions or article processing and open access charges. It will fulfil the early promise that open access would decrease costs, which is unlikely to be the case following the current policy to expect journals to flip their business model to ``gold'' publishing (a model which becomes prevalent in the UK and the EU following legal requirements).

\section{Implications for the evaluation of scientists} \label{sec4}
The competition between scientists for money, tenure and honor is necessarily competitive since these resources are scarce. We believe that this inevitable competition does not \textit{de facto} negatively affect the quality of science. 
However, the current terms of this competition definitely do (i.e. striving to secure rare publications slots in top journals as a stamp of approval). Instead, we propose a community evaluation system that removes what is an artificial construct - the rarity of scientific recognition. In our model, the quest for individual recognition does not succeed at the expenses of peers; rather it originates from open and fruitful interactions with peers. In the terms of our model, the competition for money, tenure and individual honor drives scientists to adopt a collaborative behaviour. 

This apparent paradox is not a contradiction because our processes of evaluation are global, while money, positions and prestige are delivered through local processes that involve a minority of the scientific community. For instance, it is clear that two scientists competing for the same position will hardly sing each other's praises. However, the evaluation of their work will depend mostly on what the rest of the community thinks, and in their attempt to get this positive feedback, the contenders will have had to perform openly valid and important research, write thoughtful reviews, and maintain an enlightening self-journal -- all things good for the progress of science. In the current system, even if they do not apply to the same grants, scientists working on the same topic are always a threat to each other - because the one who will publish first will mechanically shrink the value of the other\footnote{For instance, when two different labs conduct similar research and submit similar results, editors will tend to publish the article that is first submitted to them because journal prestige relies on citation counts, and science usage is such that due to anteriority it will be the one that accrues the most citations. Conversely, in the self-journal system, articles with similar results are likely be curated in pair, and therefore have similar levels of importance since the point that the curator may want to make based on them will be strengthened by the inclusion of two independent and convergent studies.}.

Therefore, we argue that the gradual adoption of the quantifiers we introduce here by institutions and funders will inject positive incentives for both quality and openness into the scientific community. The precise way that such quantifiers will be assimilated in internal processes is a choice that is political in nature an is the responsibility of those who want to evaluate researchers. It will depend on their specific goals, vision and the particularities of how best to fund science and honour scientists at the local level. If an institution can adopt the h-index, it can devise a similar index based on our quantifiers that will have a sounder scientific basis and produce positive systemic implications. Also, the modus operandi of evaluation commitees can evolve because their members can now have access to the collective intelligence of scientists who have already thought about and contextualized the articles upon which funding decisions and assignment of prestige will be based. It will be much easier and intellectually satisfying for grant reviewers who will now have access to a whole range of explicit judgements already expressed by peers. A major incentive is that, for members of evaluation committees who cannot read full articles, reading such judgements is less time-consuming, while providing a scientific safeguard and diversity they may have lacked otherwise. 

Finally, our evaluation model creates the potential for also taking into account the reviewing and curation activity of a scientist. Since these activities have an accountable influence on the course of science, they provide an impetus for many institutions (especially those struggling to compete with ``centres of excellence'') to push for a multi-dimensional approach to assessing a scientist's value, and develop policy accordingly. Just as journal-based evaluation has brought the publish-or-perish culture, our metrics can help recreate the sense of community now lacking in science. When such a different mindset is generalized, novel models of funding can also  become possible such as the decentralized approach proposed in \cite{funding}.

\section{Conclusion: towards a practical change}

We have proposed and described an evaluation system that promotes a well-defined notion of scientific value in which objective and subjective aspects are disentangled as validity and importance. The system restores a global conversation between scientists and gives control of the evolution of science back to the scientific community through open and horizontal processes, while at the same time incentivizing fruitful interactions between peers. The system also generates novel metrics that are as easy to use as the impact factor in the context of institutional evaluation or grant reviewing. It can progress in a bottom-up fashion without conflicting with current practices of publication. Since this system gives every scientist autonomous means to review and evaluate all scientific items, we consider it to be of great appeal especially to those who lack such means at present -- the junior scientists. We particularly encourage them to realize that it is their self-interest to nourish such an alternative evaluation system and reshape the power structure of tomorrow's science.

The evolution we are looking forward to is not an utopia. The science world already has a shining example of a bottom-up achievement initially powered by junior scientists: the pre-print server arXiv.org. Built in the 1990s by Paul Ginsparg, it offered a service of objective scientific value and rose out of the concrete needs of a community which supported and maintained it, despite not offering any reward in terms of institutional evaluation. Today it has become the main portal where physicists, mathematicians and computer scientists share their work in the form of preprints which are often cited more than analogous journal publications. This practice is now self-evident for these communities. 

We further propose an implementation of the ideas developed in this article in the form of SJS, an open platform for curation and peer-review, governed by an open-membership organisation of volunteer scholars\footnote{\url{www.openscholar.org.uk}}. We believe SJS could become a new multidisciplinary agora where scholars can co-validate and co-evaluate their research products while receiving, at the same time, proper recognition that can translate into both direct and immediate career benefits. Maintaining a self-journal is a rewarding, simple and risk-free activity for scientists, and debating with peers is natural in our community. We look forward to watching this academic ecosystem grow, self-organize and evolve in the years ahead.




\bibliographystyle{unsrt}

\end{document}